\documentclass[review]{elsarticle}

\usepackage{setspace}
 \usepackage{amsmath}
 \usepackage{mathtools}
 \usepackage{amssymb}
 \usepackage{graphicx}
 \usepackage{booktabs}
 \usepackage{hyperref}
 \usepackage{url}
 \usepackage{color}
 \usepackage{ulem} 

\usepackage{natbib}
\usepackage[english]{babel}

 \usepackage[dvipsnames]{xcolor}
 \usepackage[toc]{appendix}
 
 \newcommand{\tonde}[1]{\left(#1\right)}
 \newcommand{\pt}[1]{\left(#1\right)}
 \newcommand{\pq}[1]{\left[#1\right]}
 \newcommand{\pg}[1]{\left\lbrace#1\right\rbrace}
 \newcommand{\quadre}[1]{\left[#1\right]}
 
 \newcommand{\graffe}[1]{\left\{#1\right\}}

 \newcommand{\be}{\begin{equation}}
 \newcommand{\ee}{\end{equation}}
 \newcommand{\ba}{\begin{eqnarray}}
 \newcommand{\ea}{\end{eqnarray}}
 
 \newcommand{\Y}{{\mathbf{Y}}}
 \newcommand{\Yt}{{\mathbf{Y}^{\tonde{t}}}}
 \newcommand{\Ytm}{{\mathbf{Y}^{\tonde{t-1}}}}
 \newcommand{\Yijt}{{Y_{ij}^{\tonde{t}}}}
 \newcommand{\Yij}{{Y_{ij}}}
 \newcommand{\Yijtm}{{Y_{ij}^{\tonde{t - 1}}}}
 \newcommand{\Yjit}{{Y_{ji}^{\tonde{t}}}}

 \newcommand{\X}{{\mathbf{X}}}

 \newcommand{\Xijt}{{X_{ij}^{\tonde{t}}}}
 \newcommand{\Xij}{{X_{ij}}}

 \newcommand{\etm}{^{\tonde{t-1}}}
 \newcommand{\et}{^{\tonde{t}}}
 \newcommand{\uij}{_{ij}}
 \newcommand{\e}[1]{^{\tonde{#1}}}
 \newcommand{\etprime}{^{\tonde{t}\prime}}
 \newcommand{\etp}{^{\tonde{t+1}}}

 \newcommand{\allfit}{f}

 \newcommand{\bwpar}{\varphi}
 \newcommand{\ibwpar}{{\overleftarrow{\bwpar}}}
 \newcommand{\obwpar}{{\overrightarrow{\bwpar}}}
 
 \newcommand{\indfun}{\mathbb{I}}
 \newcommand{\binpar}{\theta}
 
 \newcommand{\ibinpar}{{\overleftarrow{\binpar}}}
 \newcommand{\obinpar}{{\overrightarrow{\binpar}}}
 \newcommand{\iobinpar}{\pt{\ibinpar , \obinpar }}
 \newcommand{\wpar}{\eta}
 \newcommand{\iwpar}{{\overleftarrow{\wpar}}}
 \newcommand{\owpar}{{\overrightarrow{\wpar}}}
 \newcommand{\iowpar}{\pt{\iwpar , \owpar }}

 \newcommand{\bindeg}{D}
 
 \newcommand{\ibindeg}{{\overleftarrow{\bindeg}}}
 \newcommand{\obindeg}{{\overrightarrow{\bindeg}}}
 
 \newcommand{\wdeg}{S}
 \newcommand{\iwdeg}{{\overleftarrow{\wdeg}}}
 \newcommand{\owdeg}{{\overrightarrow{\wdeg}}}

\newcommand{\sdS}{\mathcal{I}}
\newcommand{\sdw}{w}
\newcommand{\sdb}{b}
\newcommand{\sda}{a}

 \newcommand{\argmax}[1]{\underset{#1}{\operatorname{arg}\,\operatorname{max}}\;}

\makeatletter
\def\ps@pprintTitle{%
  \let\@oddhead\@empty
  \let\@evenhead\@empty
  \let\@oddfoot\@empty
  \let\@evenfoot\@oddfoot
}
\makeatother

\begin{document}
\begin{frontmatter}

\title{Score Driven Generalized Fitness Model for \\Sparse and Weighted Temporal Networks}

\author[cnr_isti_address]{Domenico Di Gangi \corref{mycorrespondingauthor}}
\author[unibo_address]{Giacomo Bormetti}
\author[sns_address,unibo_address]{Fabrizio Lillo }

\cortext[mycorrespondingauthor]{Corresponding author: digangidomenico@gmail.com}


\address[cnr_isti_address]{ISTI-CNR, via G. Moruzzi 1, 56124 Pisa}
\address[unibo_address]{Department of Mathematics, University of Bologna, Piazza di Porta San Donato 5, 40126 Bologna, Italy}
\address[sns_address]{Scuola Normale Superiore, Piazza dei Cavalieri 7, 56126, Pisa, Italy}

\begin{abstract}
   While the vast majority of the literature on models for temporal networks focuses on binary graphs, often one can associate a weight to each link. In such cases the data are better described by a weighted, or valued, network. An important well known fact is that real world weighted networks are typically sparse. We propose a novel time varying parameter model for sparse and weighted temporal networks as a combination of the fitness model, appropriately extended, and the score driven framework. We consider a zero augmented generalized linear model to handle the weights and an observation driven approach to describe time varying parameters. The result is a flexible approach where the probability of a link to exist is independent from its expected weight. This represents a crucial difference with alternative specifications proposed in the recent literature, with relevant implications for the flexibility of the model. 
    Our approach also accommodates for the dependence of the network dynamics on external variables. We present a link forecasting analysis to data describing the overnight exposures in the Euro interbank market and investigate whether the influence of EONIA rates on the interbank network dynamics has changed over time. 

\end{abstract}
\begin{keyword}
Temporal Networks\sep  Weighted Networks \sep Score Driven Models \sep Interbank Market 
\end{keyword}

\end{frontmatter}

\section{Introduction}

In the last two decades, networks, or graphs, have attracted an enormous amount of attention as an effective way of describing pairwise relations in complex systems (\cite{barabasi2002linked}).  The ever increasing abundance, and variety of graph data has motivated a great deal of applications of statistical models to graphs \citep[see, for example][for a review]{newman2010networks}. More recently, the availability of time varying networks' data has stimulated the development of models for temporal networks \citep[][]{hanneke2010discrete, sewell2015latent, giraitis2016estimating, mazzarisi2020dynamic, di2019score}.  
While the vast majority of this literature focuses on \textit{binary graphs}, i.e. graphs that are defined solely by a set of nodes and a set of links between pairs of nodes, often one can associate a weight to each link. Links' weights are typically positive, discrete or continuous, numbers and can be associated, for instance, to the strength of the relation described by each link. In standard binary descriptions, such relevant information is completely lost. For example, in a network of exposures among financial institutions the weight could be the value of the credit. In this case, a binary network would describe in the same way a link associated to an exposure of $1$ million to that associated to an exposure of $1$ billion. 
Indeed it is very common for network data to have also informative weights associated with their links. Some additional examples of the importance of weights in networks can be found, for example in: the International Trade Network \citep{leamer1995international, fagiolo2010evolution}, migration flows \citep{fagiolo2016revisiting}, scientific collaborations \citep{newman2001scientific}, transportation networks \citep{barrat2004architecture}, just to mention a few. In these cases the data are better described by a weighted, or valued, network, that can be associated with a positive, real valued matrix $\Yij \in \mathcal{R}^+ $. $\Yij$ is the weight of the  directed link from node $i$ to node $j$, and $\Yij = 0$ if the link is not present. 
Moreover, it is well known that real world networks, both binary and weighted, are very often found to be sparse, i.e. their adjacency matrices have an abundance of zero entries. That is the case, for example, of interbank networks (\cite{anand2017missing}), a class of weighted temporal networks of paramount importance, that are known to be extremely relevant to financial stability (\cite{allen2011networks,haldane2011systemic}), and have motivated the application and development of a number of statistical models for networks (see, for example, \cite{bargigli2015multiplex, mazzarisi2020dynamic} and references therein).

In spite of their relevance, networks' weights have received less attention in the literature on models for temporal networks. Indeed there are only a few models for temporal networks that take them into account. In this paper we propose a novel model for sparse and weighted temporal networks, that also accommodates for the dependency of the network dynamics on external covariates. 
{Our efforts are originally motivated by the need to properly model weighted temporal network data, describing overnight exposures in the Euro interbank market (eMID). In a previous work~\citep{di2019score}, we disregarded the information associated with the links' weight and we only focused on the temporal evolution of binary relations. We achieved that by extending the Exponential Random Graph models (ERGM), a class of statistical models for random networks, whose first and probably most famous example is the Erd{\H{o}}s-R{\'e}nyi model~\citep{erdds1959random}. In the novel framework, the ERGM parameters change over time following an observation-driven dynamics~\citep{cox1981statistical}. The relevant information for the time evolution is encoded in the filtration $\mathcal{F}_t$. It determines the update of the time varying parameters through an autoregressive process whose innovation term corresponds to the scaled score of the observation probability mass function. Such specification, known as Dynamic Conditional Score-driven (DCS) or Generalized Autoregressive Score (GAS), has been pioneered and recently introduced in the econometric literature by~\citep{creal2013generalized,harvey2013dynamic}. In this paper, we extend the well known fitness model~\citep{Holland81anexponential, PhysRevLett.89.258702,PhysRevE.78.015101, chatterjee2011random, yan2016asymptotics, yan2019statistical} for static binary networks, combining it with a simple generalized linear model to handle the weights and the DCS approach to time varying parameters. The generalization is non trivial. We need to explicitly account for the abundance of zeros that follow from the sparse nature of real world networks. We solve the issue by resorting to zero augmentation (ZA). The resulting modeling framework is very general and extremely flexible. It allows to decouple the probability of a link to exist from its expected weight -- a fact that will prove to be crucial in the forecasting exercise -- leaving us full flexibility concerning the specification of the weight distribution  and the possibility to explore the influence of external covariates on the network's dynamics.} We provide convincing Monte Carlo evidence that the score driven model is an effective filter of the latent fitness dynamics in miss-specified settings. We document a clear computational advantage, in terms of mean squared error and mean absolute difference, with respect to competitor models also in presence of external covariates and omitted variables. Interestingly, by
we exploitIn the empirical analysis, we explore from different perspectives the role played by the Euro Overnight Index Average (EONIA) rates on the dynamics of the lending relations. Consistently with similar findings discussed in the literature, we observe that lower interest rates are related with a reduction of network interconnectdness but an increase of the average liquidity flow for the loans that are present. However, the novelty of our results rely on two important facts: i) the time varying fitness model accounts explicitly for bank specific effects and thus provides a measure of the impact of EONIA rates decoupled from node specific effects; ii) our results leverage the full information available in the description of the eMID network without the need to collapse the network matrices into a single statistic, as done in \cite{akram2010interbank} and \cite{brunetti2019interconnectedness}. Finally, concerning the link and weight persistence analysis, we complement the work of \cite{hatzopoulos2015quantifying} highlighting the tendency of banks to form links whose weight is positively related with the weights at previous time steps. We call weight persistence this aspect of the dynamics of interbank relations.

The rest of the paper is organized as follows. The next section provides an overview of the main contributions proposed in the literature for the modeling of temporal weighted networks. Section~\ref{sec:SDGFM} defines a novel observation driven model for sparse and weighted temporal network, generalizing the well-known fitness model for binary networks to the sparse weighted case and leveraging the score driven approach to time varying parameters. Section~\ref{sec:num_sim_dwfm} presents an extensive Monte Carlo analysis of the score driven generalized fitness model.
Section~\ref{sec:emid_app} details the application of the new model to the eMID market. Section~\ref{sec:conclusions} draws the main conclusions and provides some ideas for future research.

\subsection{Literature Review}\label{sec:lit_rev}

An econometric model for weighted temporal networks can be described in terms of a probability distribution $P\pt{\Yijt\vert \Y\e{t-1},\dots \Y\e{1}, \X_1\e{t}, \dots, \X_{K}\e{t}}$ that describes the probability of the weight $\Yijt$ of the link between node $i$ and $j$ at time $t$ as potentially depending on previous realizations of the network and a set of contemporaneous, matrix valued, external variables $\X_1 \dots \X_K$. The external variables can be different for each link.  Even disregarding the dependency on external variables, it is immediately evident that the matrix valued nature of networks' data implies that the problem is typically very high dimensional.

Formally, observations of a weighted temporal network $\Yijt$ are no different from balanced panel data, where the cross sectional index $ij$ runs over all possible $N^2$ links. Thus, one could directly apply the non-linear panel regression methods \citep[described, for example, in][]{wooldridge2010econometric} to the sequence of positive valued matrix observations and estimate, for example, the effect of external variables. Such a direct application would nevertheless disregard the network structure and the fact that links observations are very much influenced by their association with specific nodes. Moreover, the high dimensional nature of the problem complicates the estimation and interpretation of link specific fixed effects. For this reasons, the direct approach of treating sequences of networks as standard panel data, possibly with lagged dependency, is not widespread in the literature, and restrictions to reduce the number of parameters to be estimated are used in most cases.
For example, \cite{giraitis2016estimating}, driven by system specific insights, select a set of network statistics $G_i\pt{\Y}$ and estimate the dependency of each link $\Yijt$ on $G\pt{\Ytm}$, by means of a Tobit model with few covariates, for each link. Moreover they use a local-likelihood method to estimate time varying coefficients of the regression. The censored regression used in their paper has the downside of requiring a joint modeling of the presence of a link and of its weight. While this choice allows for straightforward estimates and builds on the well known Tobit regression, it models jointly the effect that a covariate has on the probability of observing a link and the one that it has on its expected weight. In this work we propose a more flexible method that, among other things, allows to disentangle the probability of a link being present from its expected weight.

Another widespread approach is that of dynamical latent space models (\cite{kim2018review})  where a set of time varying parameters is associated to each node, and, at each time, the probability of observing a link between two nodes depends on a measure of distance between two nodes in such a latent space. Latent space models have also been considered for sparse weighted networks \citep[for example in][]{sewell2015latent}, with the aim of inferring from the network's dynamics the positions of nodes in a latent space. The resulting embedding of the nodes is typically analyzed and compared with available metadata on the nodes. While very informative, the analysis of networks' embeddings is fundamentally different from the purpose of our work, as one of main aims is to leverage the availability of data on external variables that are expected to be related with network dynamics and estimate how much the latter depends on them. 

Models that allow each one of the matrix elements $\Yijt$ to depend on each of the $\Yij\etm$ have also been considered in the literature. For example, \cite{billio2018bayesian} estimate a tensor regression (very similar to a VAR on $vec(\Y)$), with rank restrictions on the (huge) matrix of model's parameters. Differently from our work, they do not take the sparse nature of networks explicitly into account.
\cite{doi:10.1080/02664763.2017.1357684} consider a penalized logistic auto-regression model for binary networks (basically a logistic regression for each link, using all lagged matrix elements, and also products, with a lasso penalization). The same approach can in principle be extended to sparse weighted networks, and in the ZA framework.

We conclude this literature review citing some contributions that are relevant to the present work, even if they do not address directly the temporal evolution of weighted networks. Among the many papers that address the issue of modeling static sparse weighted networks, \cite{cimini2014reconstructing} consider a combination of the fitness and the gravity model of \cite{tinbergen1962shaping} to reconstruct sparse weighted financial networks. Even if they do not investigate temporal networks, their approach has proved to be very effective in modeling sparse weighted networks describing financial systems \citep[as discussed in][]{anand2015filling}.
Another interesting recent contribution is~\cite{gandy2021compound}, where the authors propose a modelling framework for weighted network data based on the compound Poisson model. They incorporate the binary fitness model as a special case but use an additional set of fitness to describe the distribution of the weights, in addition to the probability of a link being present.

Finally, we mention three recent contributions related to our work that model binary temporal networks. The first one is that of \cite{mazzarisi2020dynamic}, where the authors consider a model with time varying fitness and combine it with Discrete Auto-regressive Models \citep{jacobs1978discrete} to investigate link persistence in financial networks. Their approach is very much related to ours as it extends the binary fitness model to the temporal domain, by allowing the fitness to evolve in time. In their case the fitness follows a parameter driven dynamics. Moreover they consider a mechanism to copy links from the past, and explore the possibility of decoupling the link persistence implied by this mechanism from the probability to form new connections captured by the fitness. The second one is the recent \citep{Williams22} which introduces a non-Markovian model of binary temporal networks based on an extension of Discrete Autoregressive Models. Interestingly, authors considers also a local likelihood estimation approach to take into account non-stationarty and infer time varying parameters.  The third paper, anticipated in the Introduction, is~\citep{di2019score} where we extend the ERGMs for binary networks -- a class which includes the fitness model -- to the temporal context by allowing its parameters to evolve in time. The updating equation of the time varying parameters follows an autoregressive process whose innovation corresponds to the scaled score of the observation distribution. We show how to exploit the model as an effective filter of miss-specified dynamics and how to deal with cases where one does not have full knowledge of the likelihood function. The present paper owns a lot to the score-driven extension of the ERGM framework (SD-ERGM). Nonetheless, as it will be clear in the next section, the generalization of the static fitness model to the temporal sparse weighted case following the steps of the SD-ERGM is far from being trivial.

\section{Score Driven Generalized Fitness Model}\label{sec:SDGFM}

In this work, we propose a model for sparse and weighted temporal networks that  we refer to as \textit{score driven generalized fitness model}. Our model combines the well known \textit{fitness model} - that was first discussed in \citep{zermelo1929berechnung}, appropriately extended to a weighted zero augmented version to handle the weights, and the DCS approach to time varying parameter models \citep{creal2013generalized, harvey2013dynamic}. Before presenting the details of our model, we deem appropriate to review some preliminary concepts on the binary fitness model and the score driven framework.  
\subsection{Preliminary Concepts}
The first concept that we use in our work is the fitness model. Multiple variations of the fitness model have been considered in the literature  \citep{Holland81anexponential, PhysRevLett.89.258702,PhysRevE.78.015101, chatterjee2011random, yan2016asymptotics, yan2019statistical} and the same specification of the probability for a binary, directed or undirected, networks is known also as \textit{beta model} or \textit{configuration model}. Hereafter we will define the fitness model as a model for binary networks where, in the case of directed networks,  two parameters are assigned to each node $i$, the in-fitness $\ibinpar_i $ and the out-fitness $ \obinpar_i$, that describe the tendency of each node to form incoming and outgoing connections, respectively. Because the occurrence of a link is very likely influenced by both intrinsic and external factors, we focus in the following on a specification that allows for the links' probability to depend also on the external, possibly link specific, covariates \citep{yan2019statistical}. Hence  the probability of a link to exist is described by 
\begin{equation*}
    p_{ij}  = \frac{1}{1+ e^{ - ( \ibinpar_i +  \obinpar_j + X_{ij} \beta )}},
\end{equation*}
where $\beta$ is the coefficient tied to the information of the, link specific, covariate $\X$.
This model is very convenient because it allows for a parsimonious modeling of the matrix associated with a network, via node specific fitness. 

The second key ingredient of our approach is the DCS framework for time varying parameter models, introduced by~\cite{creal2013generalized} and~\cite{harvey2013dynamic}. In order to review it, let us consider a sequence of observations $\graffe{z\et}_{t=1}^T$, where each $z\et \in\mathbb{R}^M$,  and a conditional probability density $P\tonde{z\et\vert \allfit\et}$, that depends on a vector of time varying parameters $\allfit\et \in \mathbb{R}^K$. Defining the score as $\nabla\et = \frac{\partial \log{P\tonde{z\et\vert \allfit\et}}}{\partial \allfit\etprime}$, a score-driven model assumes that the time evolution of $\allfit\et$ is ruled by the recursive relation
\be  \label{eq:gasupdaterule}
\allfit\etp = \sdw +\boldsymbol{\sdb} \allfit\et + \boldsymbol{\sda} \boldsymbol{\sdS\et} \nabla\et , \\
\ee 
where $\sdw$, $\boldsymbol{\sda}$ and $\boldsymbol{\sdb}$ are static parameters, $\sdw$ being a $K$ dimensional vector and $\boldsymbol{\sda}$ and $\boldsymbol{\sdb}$  $K\times K$ matrices.  $\boldsymbol{\sdS\et}$ is a $K\times K$ scaling matrix, that is often chosen to be the inverse of the square root of the Fisher information matrix associated with $P\tonde{y\et\vert \allfit\et}$, i.e. $\boldsymbol{\sdS\et} = \mathbb{E}\quadre{\frac{\partial \log{P\tonde{y\et\vert \allfit\et}}}{\partial \allfit\etprime} \frac{\partial \log{P\tonde{y\et\vert \allfit\et}}}{\partial \allfit\etprime}^\prime }^{-\frac{1}{2}}$. However, this is not the only possible specification and different choices for the scaling are discussed in~\cite{creal2013generalized}. 
A score driven model can be regarded both as \textit{data generating process} (DGP) or as a filter of an unknown dynamics. In both cases, the most important feature of \eqref{eq:gasupdaterule} is the role of the score as the driver of the dynamics of $f\et$. The structure of the conditional observation density determines the score, from which the dependence of $f\etp$ on the vector of observations $y\et$ follows. When the model is viewed as a DGP, the update results in a stochastic dynamics thanks to the random occurrence of $z\et$. When the score-driven recursion is regarded as a filter, the update rule in \eqref{eq:gasupdaterule} is used to obtain a sequence of filtered  $\pg{\hat{f}\et}_{t=1}^T$. In this setting, the static parameters are estimated maximizing the log-likelihood of the whole sequence of observations \citep[for a detailed discussion, see ][]{harvey2013dynamic, blasques2014maximum}, 
\begin{equation*}
\pt{\hat{\sdw},\hat{\boldsymbol{\sdb}},\hat{\boldsymbol{\sda}}}  = \argmax{\pt{\sdw,\boldsymbol{\sdb},\boldsymbol{\sda}}} \sum_{t=1}^T \log{P\pt{ z\et  \vert f\et \pt{\sdw,\boldsymbol{\sdb},\boldsymbol{\sda},\pg{z^{\pt{t^\prime}}}_{t^\prime=1}^{t-1}} }}.
\end{equation*}

Score driven models have seen an explosion of interest in recent years due to their flexibility and ease of estimation. Indeed, many state of the art wildly popular econometric models turn out to belong to the family of score driven models. Examples are the Generalized Autoregressive Conditional Heteroskedasticity (GARCH) model of \cite{BOLLERSLEV1986307}, the Exponential GARCH model of \cite{nelson1991conditional}, the Autoregressive Conditional Duration model of \cite{engle1998autoregressive}, and the Multiplicative Error Model of \cite{engle2002new}.  Moreover, there are motivations, originating in information theory, for the optimality of the score-driven updating rule \cite{blasques2015information}. The introduction of this framework in its full generality has been investigated from a theoretical point of view \citep[][]{blasques2014maximum, blasques2015information, blasques2017finite, harvey2020modeling} and opened the way to applications in various contexts~\footnote{Please refer to \url{http://www.gasmodel.com/index.htm} for the updated collection of papers dealing with GAS models.}. 

\subsection{Definition of the Model}\label{sec:new_mod_def}

In order to introduce the \textit{score driven weighted fitness model}, let us describe a sequence of networks with a set of random variables $\Yijt$, one for each link. We propose to use Zero Augmentation to model separately the probability of observing a link, $\Yijt > 0$, and the probability to observe a specific weight $\Yijt$. We prefer Zero Augmentation over censoring, as we believe the former to be much more flexible in the context of network data. With this choice the probability distribution for the link $ij$ is
\begin{equation}\label{eq:zero_aug_w_nets_pdf}
P\tonde{\Yijt  = y} =  \left\lbrace\begin{array}{ll} 
	1-p\uij\et    \quad &for \quad y=0 \\
	 & \\
	p\uij\et  g\uij\et\tonde{y}  \quad &for \quad y>0 \, . \end{array}\right.  \\
\end{equation}
where $ g\uij\et\tonde{y}$ is the distribution for the positive continuous weight for link $ij$, conditional on the presence of the link.

We then model the binary temporal network and its weights by means of time varying fitness, allowing also for the dependency on external covariates $X_{ij}$. In our model the probability of a link to exist is described by
\begin{equation}\label{eq:gen_fit_bin_prob}
    p_{ij}\et  = \frac{1}{1+ e^{ - ( \ibinpar_i\et +  \obinpar_j\et + X\et_{{ij}} \beta_{bin} )}},
\end{equation}
where, for simplicity we consider only one external covariate $\Xijt$ with its own associated parameter $\beta_{bin}$, while in general nothing prevents us to have multiple covariates. With this choice, the log-likelihood for the single observation $\pt{\Y\et}$ in \eqref{eq:zero_aug_w_nets_pdf} is
\begin{equation*}
\begin{aligned}
\log{P\tonde{\Y\et \vert \ibinpar\et, \obinpar\et, \beta_{bin}, \beta_{w}, \X\et}}  &= \sum\uij \pt{\indfun\pt{Y_{ij}\et} - 1} \pt{\ibinpar_i\et + \obinpar_j\et + \beta_{bin} X\et_{ij}} \\ 
&- \log{\tonde{1 + e^{-\pt{\ibinpar_i\et + \obinpar_j\et + \beta_{bin} X\et_{ij}}} }} + \indfun\pt{Y_{ij}\et} \log{g\uij\et} \\
&=\sum_i \ibindeg_i \ibinpar_i\et + \obindeg_i \obinpar_i\et   \\ 
& + \sum\uij   \indfun\pt{Y_{ij}\et} \beta_{bin} X\et_{ij} + \ibinpar_i\et + \obinpar_j\et + \beta_{bin} X\et_{ij}\\
&- \log{\tonde{1 + e^{-\pt{\ibinpar_i\et + \obinpar_j\et + \beta_{bin} X\et_{ij}}} }} + \indfun\pt{Y_{ij}\et} \log{g\uij\et},
\end{aligned}
\end{equation*}
where the indicator function $\indfun$ is zero if its argument is less or equal than zero and one otherwise. We denote the in and out degrees of vertex $i$ at time $t$ respectively as $\ibindeg_i = \sum_j\indfun\pt{\Yijt}$ and $\obindeg_i = \sum_j\indfun\pt{\Yjit}$.
As in the standard fitness model for binary networks, the time varying \textit{binary fitness} parameters $\pt{\ibinpar\et, \obinpar\et}$ describe the tendency of nodes at time $t$ to form links, that are not explained by the external covariate $\X\et$. 
In order to model the weights of the observed links, we consider a generalized version of the fitness model where we associate to each node $i$, at time $t$, also the parameters $ \iwpar_i\et, \owpar_i\et$, that we call \textit{weighted fitness}. They describe the propensity of a node to have more or less heavy weights in incoming and outgoing links respectively, and are related to the distribution of the weights of present links $g_{ij}\et$ by 
\begin{equation}\label{eq:gen_fit_cond_exp}
    \mathbb{E}\pq{\Yij\et\vert \Yijt>0 } = e^{\tonde{\iwpar_{i}\et + \owpar_{j}\et + X_{ij}\et \beta_{w}}},
\end{equation}
where we considered the dependency on a single external covariate $\Xijt$ and indicated the associated regression coefficient with $\beta_{w}$ to distinguish it from the coefficient for the binary part in \eqref{eq:gen_fit_bin_prob}.
This choice of linking the weighted fitness to the mean of the distribution $g$ provides dynamics and heterogeneity only to one parameter of the conditional distribution, as shown in the following with a concrete example. The score driven weighted fitness model can be defined for a generic distribution $g$, for both continuous and discrete data, as we discuss in Appendix~\ref{sec:weight_distrib_SI}. Nevertheless in the following, for concreteness,  we will focus on the gamma distribution to model links' weights
\begin{equation*}
g\uij\tonde{y} =  \frac{\pt{\mu_{ij}}^{-\sigma} y\e{\sigma-1}}{\Gamma\pt{\sigma}}  e^{-{\frac{y}{\mu_{ij}}}}\,,
\end{equation*}
where $\sigma$ is a positive constant. Given a sequence of observed  weighted adjacency matrices $\graffe{\Yt}_{t=1}^T$, we denote by $\allfit\et$ a $K$ dimensional vector, where $K=4\times N$, containing all the time varying fitness parameters $\ibinpar\et, \obinpar\et,  \iwpar\et, \owpar\et$. With this notation, the model's distribution takes the following form
\begin{equation}\label{eq:gen_gamma_fit_mod}
P\tonde{\Yijt  = y \vert \allfit\et , \beta_{bin}, \beta_{w}, \sigma } =  \left\lbrace\begin{array}{ll} 
	\frac{e^{ - ( \ibinpar_i\et +  \obinpar_j\et + X\et_{{ij}} \beta_{bin} )}}{1+ e^{ - ( \ibinpar_i\et +  \obinpar_j\et + X\et_{{ij}} \beta_{bin} )}}  \quad & for \quad y=0 \\
	\\ 
	\frac{ \pt{\mu_{ij}\et}^{-\sigma} \Gamma\pt{\sigma}^{-1}  }{1+ e^{ - ( \ibinpar_i\et +  \obinpar_j\et + X\et_{{ij}} \beta_{bin} )}} y\e{\sigma-1}  e^{-{\frac{y}{\mu_{ij}\et}}}   \quad &for \quad y>0 \, . \end{array}\right.
\end{equation}
with 
$$\mu_{ij}\et = \sigma^{-1} e^{\tonde{\iwpar_{i}\et + \owpar_{j}\et + X_{ij}\et \beta_{w}}}.$$

We let the fitness, both binary and weighted, evolve in time, following the score-driven recursive update rule in \eqref{eq:gasupdaterule}, that in this case takes the form
\be  \label{eq:fit_sd_update}
\allfit\etp = \sdw +\sdb \allfit\et + \sda \boldsymbol{\sdS\et}  \frac{\partial \log{P\tonde{\Yt\vert \allfit\et}}}{\partial \allfit\etprime} , \\
\ee 
where $\sdw$, $\sda$ and $\sdb$ are three $K$ dimensional vectors of static parameters~\footnote{Hence, in our definition we have three static parameters for each time varying fitness. While in the general formulation of score driven models  \citep[as given, for example, in][]{creal2013generalized}, the parameters $\sdb$ and $\sda$ are defined as matrices, it is very common to impose some restriction on them in order to limit the number of parameters to be estimated, as we do here.}. $\boldsymbol{\sdS\et}$ is a $K\times K$ scaling matrix. Hence, conditionally on the value of the parameters $\allfit\et$ at time $t$ and the observed adjacency matrix $\Y\et$, the parameters at time $t+1$ are deterministic. The element $k$ of the score for the in and out binary fitness takes the following form
\begin{align}\label{eq:score_fit_bin}
    \frac{\partial \log{P\tonde{\Y \vert \allfit\et, \beta_{bin}, \beta_{w}, \X\et}} }{\partial \ibinpar\etprime_k} &=  \ibindeg_k  - \sum_{j} \frac{1}{1+ e^{ - ( \ibinpar_k\et +  \obinpar_j\et + X\et_{{kj}} \beta_{bin} )}}   \nonumber \\
    \nonumber \\
     \frac{\partial \log{P\tonde{\Y \vert \allfit\et, \beta_{bin}, \beta_{w}, \X\et}} }{\partial \obinpar\etprime_k} &= \obindeg_k - \sum_{i}  \frac{1}{1+ e^{ - ( \ibinpar_i\et +  \obinpar_k\et + X\et_{{ik}} \beta_{bin} )}}   
\end{align}
and does not depend on the choice of $g$. The element $k$ of the score for the weighted in and out fitness are
\begin{align}\label{eq:score_fit_w}
    \frac{\partial \log{P\tonde{\Y \vert \allfit\et, \beta_{bin}, \beta_{w}, \X\et}}}{\partial \iwpar\etprime_k} &= \sum_{j} \indfun\pt{Y_{kj}\et} \frac{\partial \log g_{kj}}{\partial \iwpar_k} =  \sum_j   \indfun\pt{Y_{kj}\et}  \pt{\frac{Y\et_{kj}}{\mu\et_{kj}} - \sigma}  \nonumber \\
    \nonumber \\
    \frac{\partial\log{P\tonde{\Y \vert \allfit\et, \beta_{bin}, \beta_{w}, \X\et}}}{\partial \owpar\etprime_k} &= \sum_{i} \indfun\pt{Y_{ik}\et} \frac{\partial \log g_{ik}}{\partial \owpar_k} =  \sum_i   \indfun\pt{Y_{ik}\et} \pt{\frac{Y\et_{ik}}{\mu\et_{ik}} - \sigma} .
\end{align}
In \eqref{eq:score_fit_bin} and \eqref{eq:score_fit_w}, as scaling matrix  we use the Hessian of the log-likelihood. 

As any score driven model, our model can be regarded both as a DGP or as a filter of a misspecified dynamics. When used as a DGP, it describes a stochastic dynamics because, at each time $t$, the adjacency matrix is not known in advance. It is randomly sampled from $P\tonde{\Yt\vert \allfit\et}$ and then used to compute the score that, as a consequence, becomes itself stochastic.
When the sequence of networks $\graffe{\Yt}_{t=1}^T$ is observed, and the model is applied as a filter of the time varying parameters, the static parameters $\tonde{\sdw,\sdb,\sda}$, that best fit the data, can be estimated maximizing the log-likelihood of the whole time series. 

Taking into account that each network $\Yt$ is independent from all the others \textit{conditionally} on the value of $\allfit\et$, the log-likelihood can be written as
\be\label{eq:sd_fit_mod_likelihood}
\log{P\tonde{\graffe{\Yt}_{t=1}^T \vert w,\sdb, \sda}} = \sum_{t=1}^T \log{P\tonde{ \Yt  \vert \allfit\et\tonde{w, \sdb, \sda,\graffe{\Y^{\tonde{t^\prime}}}_{t^\prime=1}^{t-1}} }}.
\ee
Then the filtered time varying fitness $\widehat{\allfit}\et$ are obtained by an iterative application of \eqref{eq:fit_sd_update} using as parameter values the maximum likelihood estimates (MLEs).

For ease of exposition, so far we introduced a baseline version of our model where the probabilities depend on a single external covariate through the scalar $\beta_{bin}$ and $\beta_w$. This implies uniform, i.e. equal across all links, dependency on the covariates. In order to explore node specific dependency on the external variables, we could easily consider the specification  $X_{ij} \pt{\beta_{bin_i} + \beta_{bin_j}}$, where we associate two covariate coefficients to each node, one for outgoing links and one for incoming links.

\subsubsection{Poisson Distribution for Discrete Weights and Inference from Partial Information }

As mentioned before, our method is not restricted to a specific distribution for the weights and we can easily consider a different specification for $g_{ij}$ in \eqref{eq:zero_aug_w_nets_pdf}, instead of the gamma distribution. For example, in some settings it might be more appropriate to model weights as discrete variables. In this section we show how we can describe discrete weights by means of the Poisson distribution, defined as follows
\begin{equation*}
g\uij\tonde{y} =  \frac{ \mu_{ij}^{y} e^{-\mu_{ij}} }{ y!}\,~~~~~y\in {\mathbb N}.
\end{equation*}
If we consider a specification without external covariates, hence where $\mu_{ij}\et = \sigma^{-1} e^{\tonde{\iwpar_{i}\et + \owpar_{j}\et}}$, the log-likelihood takes the following form
\begin{equation*}
\begin{aligned}
\log{P\tonde{\Y\et \vert \allfit\et}} &=\sum_i \ibindeg_i\et \ibinpar_i\et + \obindeg_i\et \obinpar_i\et + \sum\uij   \ibinpar_i\et + \obinpar_j\et - \log{\tonde{1 + e^{-\pt{\ibinpar_i\et + \obinpar_j\et }} }} \\
& + \sum\uij  \indfun\pt{\Yijt} \log{g\uij\et}\\
&=\sum_i \ibindeg_i\et \ibinpar_i\et + \obindeg_i\et \obinpar_i\et + \sum\uij   \ibinpar_i\et + \obinpar_j\et - \log{\tonde{1 + e^{-\pt{\ibinpar_i\et + \obinpar_j\et }} }} \\
&+  \sum_i \pt{\iwdeg_i\et \iwpar_i\et + \owdeg_i\et \owpar_i\et - \ibindeg_i\et e^{\iwpar_i\et} - \obindeg_i\et e^{\owpar_i\et}} -  \log\pt{\Yijt !},
\end{aligned}
\end{equation*}
where we defined the in and out strengths as $\iwdeg_i\et = \sum_j\Yijt$ and $\owdeg_i\et = \sum_j\Yjit$, respectively.
Interestingly the log-likelihood of the model based on the Poisson distribution depends only on the degrees $\pt{\ibindeg\et, \obindeg\et}$ and strengths $\pt{\iwdeg\et, \owdeg\et}$ and, apart from the last not essential additive term, does not depend on the full matrix $\Yt$. The score of $\iobinpar$ does not depend on the distribution of the weights and reads as in \eqref{eq:score_fit_bin}, while the element $k$ of the score for the weighted in and out fitness is
\begin{align*}
    \frac{\partial \log{P\tonde{\Y \vert \allfit\et}}}{\partial \iwpar\etprime_k} &=  \iwdeg_k\et - \ibindeg_k\et e^{\iwpar_k\et} \nonumber \\
    \nonumber \\
    \frac{\partial\log{P\tonde{\Y \vert \allfit\et}}}{\partial \owpar\etprime_k} &=  \owdeg_k\et - \obindeg_k\et e^{\owpar_k\et} .
\end{align*}
Hence we could use the score driven update and estimate the model only relying on the sequences of strengths and degrees, ignoring the full matrix $\Yt$. This has interesting implications as it allows for a parsimonious implementation of the inference procedure. Indeed, the observations can be compressed from a sequence of $T$ matrices each with $N\pt{N-1}$ elements~\footnote{Not considering the diagonal elements that would describe self-loops.} to only $4$ sequences of vectors each of length $N$, for a total of $4N$ elements. More importantly, it is not uncommon in practice not to have access to the full adjacency matrix, but only to the degree and strength sequences. This circumstance is so relevant that has stimulated an important stream of literature on the reconstruction of networks from partial information \citep{Mis11, anand2015filling, anand2017missing, gandy2017adjustable}. We do not further investigate this aspect in this paper but leave it for future research.

\section{Numerical Simulations in Misspecified Settings}\label{sec:num_sim_dwfm}

In this section we discuss the results of extensive numerical simulations~\footnote{The python code used for the simulations is available at  \url{https://github.com/domenicodigangi/DynWGraphsPaper}} that we run to evaluate the performance of the score driven weighted fitness model as a misspecified filter, i.e. when the true DGP of the simulated data is not the same as the score driven update rule. This is the typical situation in practical applications, where the true DGP is unknown. 

When modeling a sequence of observed weighted networks with time varying fitness as in Eq. \eqref{eq:gen_gamma_fit_mod}, we compare the score driven update rule to two simple alternative filters. The first one is to specify the fitness, both binary and weighted, as static. This amounts to consider a zero augmented generalized linear model, accounting for node specific effects by means of the fitness. The probability of observing a link is that of the standard fitness model  \eqref{eq:gen_fit_bin_prob}, where the parameters $\iobinpar$ do not change for all time steps $T=1,\dots, T$, while the distribution of the weights depends on the constant weighted fitness $\iowpar$ such  that 
\begin{equation*}\label{eq:gen_fit_cond_exp_const}
    \mathbb{E}\pq{\Yij\et\vert \Yijt>0 } = e^{\tonde{\iwpar_{i} + \owpar_{j} + X_{ij}\et \beta_{w}}}\,.
\end{equation*}
The second alternative consists in estimating a static fitness model for each observed snapshot, i.e. at each time $t$. This procedure results in a sequence of single snapshot estimates for the fitness, where clearly the number of parameters to be estimated grows with the number of networks observed. This sequence of single snapshot estimates provides a filter for the time varying fitness, and has already been discussed, for the binary case, in \cite{di2019score}.

In the rest of this section we compare the score driven weighted fitness model, when appropriate, with these two alternatives in three numerical experiments. 

\subsection{Filtering Time Varying Fitness}
In our first numerical experiment we focus on filtering the fitness that evolves in time following a DGP different from the score driven dynamics. 
Specifically, we assess how accurately the score driven methodology allows us to filter the fitness when no external covariates are present, and compare it with the sequence of single snapshot MLEs. Since the focus here in on filtering time varying fitness, we do not consider the alternative version with constant fitness.
In \cite{di2019score} we already carried out a similar comparison for the binary part of the model and we showed that, for the binary part, the sequence of cross sectional estimates clearly under-performs the score driven fitness model, both in numerical simulations and empirical applications. Here we repeat a similar exercise considering also the weighted fitness.

\begin{figure}[h!]
\centering
\includegraphics[width=0.49\linewidth 	]{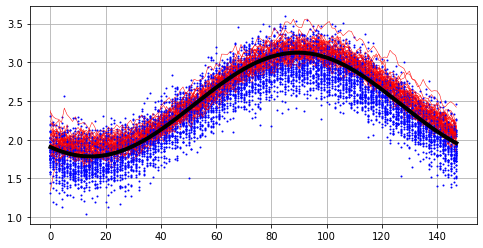}
\caption{Filtered paths for one of the time varying weighted fitness parameters. The path from the true DGP is in black. The blue dots are the single snapshot estimates, and the red lines the paths filtered with the score driven generalized fitness model.}
\label{fig:dgp_sin}
\end{figure}

Specifically, we analyze sequences of networks sampled from Eq. \eqref{eq:gen_gamma_fit_mod} where each fitness, both binary and weighted, evolves according to an auto-regressive process of order one, $\allfit_i\etp = b_0 + b_1 \allfit_i\et + \epsilon\et $ where $\epsilon \sim N\pt{0, 0.1}, \; b_1 = 0.98$ and $b_0$ is chosen for each parameter in order to sample networks that resemble a real world realization.  We also consider one example of a deterministic DGP where the fitness evolve in time following a sin function, as shown in Figure \ref{fig:dgp_sin}. In practice, in order to obtain realistic parameters values for the DGPs, we first estimate the fitness models, both binary and weighted on the first observation available for the temporal interbank network data that we describe in Section \ref{sec:app_eonia}~\footnote{The numerical values for the fitness used are available at \url{https://github.com/domenicodigangi/dynwgraphs}. We have checked that similar results hold true using different numerical values for the fitness parameters.}. 
Then, we sample time series of $150$ observations from the AR(1) processes for each fitness independently. We use the simulated fitness paths as parameters to sample from  the model specification in Eq. \eqref{eq:gen_gamma_fit_mod}. Using only the sequence of sampled matrices as input, with the two methods we filter the time varying evolution of the fitness. For each fitness we compute the  mean squared error (MSE) over the time series and then average it across nodes, keeping separate the MSE for binary and weighted fitness. Finally, we repeat the sampling and filtering procedure $50$ times and average the obtained MSE across the repetitions.
\begin{table}[ht]
\centering
\begin{tabular}{l|cc|cc}
	\rule{0pt}{12pt} DGP  & \multicolumn{2}{c}{AR(1)} & \multicolumn{2}{c}{SIN} \\
    \hline
	\rule{0pt}{12pt}  Filter  & SS & SD & SS  & SD \\
	\hline
	\hline
	\rule{0pt}{12pt}  Avg. MSE $\iobinpar$  & 0.36  & 0.11 & 0.66  & 0.04 \\
	\hline
	\rule{0pt}{12pt}  Avg. MSE $\iowpar$  & 0.54 & 0.25 & 0.6  & 0.18 \\
\end{tabular}
\caption{Results from the first experiment: MSE of the filtered fitness averaged across all nodes, for both the single snapshot (SS) and score driven (SD) filters.} \label{tab:experiment_1}  
\end{table} 
The results, showed in Table \ref{tab:experiment_1}, confirm that, also for the weighted fitness, the score driven update rule is clearly a better choice in filtering misspecified paths for the time varying fitness. 

Having established that as a misspecified filter of the fitness, the sequence of single snapshot estimates under-performs the score driven alternative, also for the weighted fitness, let us mention a second limitation of the single snapshot approach. In the usual empirical setting, only one network realization is available at each time step. In such a situation, we might not be able to jointly estimate the sequence of single snapshots estimates and the coefficients $\beta_{bin}$, or $\beta_{w}$, due to the low number of observations per parameter. This is the case, for instance, if we consider as covariate a variable that is uniform across all links i.e. $X\et_{ij} = x\et$ for all $i, j$, a case that we consider in the empirical application of Section \ref{sec:app_eonia}. An identification issue arises for the single snapshot estimates for both the binary and weighted parameters. In fact, the probability of the sequence of observations, given the sequence of weighted fitness, remains unchanged under the following transformation
\begin{align*}
\iwpar_i\et &\rightarrow \iwpar_i\et + c_1 x\et, \quad \forall i \nonumber\\ 
\owpar_i\et &\rightarrow \owpar_i\et + c_2 x\et, \quad \forall i \nonumber \\ 
\beta_w &\rightarrow \beta_w + c_3 , 
\end{align*}
for any choice of $(c_1, c_2, c_3)$ such that $c_1 + c_2 + c_3 = 0$,
since for each $t$ such a transformation does not change the sum $\iwpar\et_i + \owpar\et_j + \beta_w x\et$\footnote{The need to fix an identification condition in fitness models, even without external covariates, is well known as we discuss in Appendix  \ref{sec:identification_SI}. }. 
Hence, in the model specification with a uniform external covariate, we cannot identify the parameter of interest $\beta_w$ and this prevents us to use sequences of single snapshot estimators. With a simple change of notation we can see that the same issue arises for the binary parameters. We point out that the model with static fitness, that we use for comparison, and our score driven version, do not suffer from this identification issue, as in both cases the number of static parameters to be estimated does not increase with the number of time steps observed. For instance, in the score driven model, the sequence of time varying parameters $\pt{\iwpar\et, \owpar\et}$ for $t=1, \dots, T$, is not estimated directly but follows from the score driven update rule \eqref{eq:fit_sd_update} that is uniquely identified given the sequence of observations and the static parameters $\tonde{\sdw,\sdb,\sda}$. For this reason, in the rest of this section, we compare the score driven weighted fitness model only with the alternative having constant fitness. 

\subsection{External Covariates and Omitted Variables Misspecification}\label{sec:num_sim_dwfm_unobs}

 We present here two numerical experiments to assess how effective the score driven weighted fitness model can be in estimating the dependency of the network dynamics on external covariates. We show that, in synthetically generated datasets, introducing the binary and weighted time varying fitness reduces errors due to unobserved variables. 

Our second experiment is designed to show that the score driven model properly describes the effect of external covariates, even when the time varying fitness is generated by a misspecified DGP. Moreover it highlights the importance of taking into account time varying node specific effects. In fact, assuming the fitness to be constant, when they are actually time varying, results in poor estimates of the dependency on external covariates. 
In order to show that, we consider samples from the model in Eq. \eqref{eq:gen_gamma_fit_mod}, where we let the fitness evolve with the AR(1) DGP of the previous experiment. We also assume that the network dynamics depends on the realization of two independent, predetermined, external variables, $\X_{bin}$ and $\X_w$. The first covariate enters the binary part of the DGP
$$
p_{ij}\et  = \frac{1}{1+ e^{ - ( \ibinpar_i\et +  \obinpar_j\et + X\et_{bin\uij} \beta_{bin} )}},
$$
and the second one influences the expected weights in \ref{eq:gen_gamma_fit_mod} as follows
$$
\mu_{ij}\et = \sigma^{-1} e^{\tonde{\iwpar_{i} + \owpar_{j} + X_{w\uij}\et \beta_{w}}}.
$$
We fix $\beta_{bin} = 1$, $\beta_{w} = 1$ and consider two possible specifications for the DGP of the synthetic external covariates. In the first one, both external variables are scalar, $X\et_{bin_{ij}} = x_{bin}\et$ and  $ X\et_{w_{ij}} = x_{w}\et$ for all $i, j$, and follow an AR(1) process equal to the one followed by the fitness in the previous example~\footnote{We set the parameter $b_0$ so that the unconditional mean of the AR(1) process is equal to $1$.}. In the second specification, we set $X\et_{bin_{ij}} = \indfun\pt{\Yijtm} $ and  $ X\et_{w_{ij}} =  \indfun\pt{\Yijtm}  \log\pt{\Yijtm}$ for all $i, j$. The latter DGP, due to the explicit dependency of the network at time $t$ from its realization at previous time $t-1$, simulates a temporal network with link persistence, i.e. a higher probability of observing at time $t$ a link if this was present at time $t-1$ and weight persistence, i.e. the weight at time $t$, if the link is present, is affected by its weight at time $t-1$.

We sample 50 sequences of networks, and external covariates, each of 150 time steps. 
For each sampled sequence, we estimate the score driven weighted fitness model maximizing the likelihood of the static parameters. We then compute the MSE between simulated and estimated values of $\beta_{bin}$ and $\beta_w$ across the sample. In order to test the effect of neglecting time varying effects,  we repeat the same procedure for a version of the model with static fitness and report the results in Table \eqref{tab:experiment_2}. 
\begin{table}[ht]
\centering
\begin{tabular}{l|cc|cc}
   DGP &  \multicolumn{2}{c|}{\rule{0pt}{15pt} AR(1) Scalar External Regressors } & \multicolumn{2}{c}{\rule{0pt}{15pt}  Persistent links and weights }   \\ 
	\hline
	Filter \rule{0pt}{15pt}  &    Constant Fitness   &   SD Fitness  &    Constant Fitness   &   SD Fitness \\
	\hline
    \hline
	\rule{0pt}{12pt}  MSE $\beta_{bin}$  & 0.14  & 0.06  & 0.23  & 0.02 \\
	\hline
	\rule{0pt}{12pt}  MSE $\beta_w$  & 0.34  & 0.02   & 0.18  & 0.015 \\
\end{tabular}
\caption{Results for the second experiment: MSE of the estimated regression coefficients (average over 50 samples). The fitness dynamics follow an AR(1) process. Two columns on the left: external regressors are scalar and follow two AR(1) processes. Two columns on the right: regressors induce a persistent dynamics in both links and weights.} \label{tab:experiment_2}  
\end{table} 
It emerges clearly that not taking into account the dynamics of the fitness can severely deteriorate the estimation of the effect of external covariates.

In order to motivate our third and last numerical experiment, let us mention that the topic of estimation errors due to omitted variables has been discussed widely in the econometric literature \citep[][]{greene2000econometric, barreto2006introductory, wooldridge2010econometric}. The standard approach to mitigate it consists in using control variables. This approach has some known downsides \citep[refer for instance to][]{griliches1977estimating, yatchew1985specification, clarke2005phantom} and, most importantly, it is not always viable since the data on appropriate controls might not be available. Indeed, this is very common for financial networks, as the one that we consider in Section \ref{sec:app_eonia}, where the variables that one would like to use as controls are likely to be privacy protected and often unavailable to researchers. Considering, for example the case of interbank networks, we could very well expect the current leverage of a bank to have a significant influence on the decision to borrow or lend, i.e. create interbank links. Nevertheless it is very unlikely for this information to be available, at the frequency required. This issue is even more frequent when the datasets are anonymous, and the identity of the nodes is not known. For these reasons, 
in the third experiment we show that allowing the fitness to vary in time is extremely beneficial to mitigate the errors due to omitted variables, at least in the context of controlled numerical simulations.  We assume the network dynamics, both of the links and of the weights, to be determined by two external covariates. We then assume that, for whatever reason, only one of them is observed. Specifically the considered DGP is similar to the one in Eq. \eqref{eq:gen_gamma_fit_mod} but with parameters defined as
$$
p_{ij}\et  = \frac{1}{1+ e^{ - ( \beta_{1,bin} x\et_{1}  + \beta_{2,bin} x\et_{2} )}},
$$
and
$$
\mu_{ij}\et = \sigma^{-1} e^{\tonde{ \beta_{1,w} x_{1}\et +  \beta_{2, w} x_{2}\et }}.
$$
We assume that $x_2\et$ is not observed and assess the effect of omitting it when estimating the coefficients $\beta_{bin_1}$ and $\beta_{w_1}$. We show that the score driven fitness compensates the impact of the neglected variable on the estimates. The external variables $x_{1}\et$ and $x_{2}\et$ are independent and follow an AR(1) model with high persistence, $b_1 = 0.98$. 
\begin{table}[ht]
\centering
\begin{tabular}{l|ccc}
	  DGP &  \multicolumn{3}{c}{\rule{0pt}{15pt} No Fitness and two regressors} \\ 
	  \hline
	Filter  \rule{0pt}{15pt}  &   No Fitness &  Constant Fitness & SD Fitness  \\
	\hline
	\hline
	\rule{0pt}{12pt}  MSE $\beta_{bin_1}$ & 5.66  & 0.53 & 0.01  \\
 	\hline
	\rule{0pt}{12pt}  MSE $\beta_{w_1}$ & 133  & 0.39 & 0.08 \\
 	\hline
\end{tabular}
\caption{Misspecified filtering of a DGP with two covariates and no fitness. First column: MSE for the estimates of $\beta_{bin_1}$ and $\beta_{w_1}$, when no fitness is used (average over 50 samples). Second column: MSE for estimates with constant fitness. Third column: MSE using a model with score driven fitness.} \label{tab:experiment_3}  
\end{table} 
We sample the DGP and estimate the coefficients $50$ times for a sequence of networks $150$ time step long. We then compare the MSE for the estimates of parameters $\beta_{1,bin}$ and $\beta_{1,w}$, obtained using three different specifications of the model in \eqref{eq:gen_gamma_fit_mod}: one without node heterogeneity, hence no fitness, where the probability of observing a link depends only on the observed covariate. The second is a model  with constant fitness a the model with score driven fitness and the third uses the observed external covariate. From the results that we report in Table \eqref{tab:experiment_3},  it follows that considering a model with time varying fitness is extremely beneficial when the DGP is misspecified and does not feature node specific effects. We believe that the results presented in this section strongly support the choice to describe temporal weighted networks by means of time varying fitness. Moreover, they give us clear insights to interpret the results of Section \ref{sec:emid_app}, where we find a clear advantage, in terms of goodness of fit, in  using the score driven weighted fitness model to describe the empirical data. 
 
\section{Link and Weight Dynamics in the Italian e-MID}\label{sec:emid_app}
We apply our model to the interbank overnight loans market described as a temporal weighted network. Interbank markets are an important point of encounter for banks' supply and demand of extra liquidity, and have received much attention in the literature \citep[see][ for a review]{green2016overnight}. In particular, e-MID has been investigated in many papers
\citep[see, for example][and references therein]{iori2008network,finger2013network,mazzarisi2020dynamic,barucca2018organization}. We use data from the e-MID, a market where banks can extend loans to one another for a specified term and/or collateral. Our dataset contains the list of all credit transactions in each day from June 6, 2009 to February 27, 2015. In our analysis, we investigate the interbank network of overnight loans, aggregated weekly. The standard approach in the literature to model temporal interbank networks is to disregard the size of the exposures and consider only the presence or absence of links, i.e. to consider only the binary network. Thanks to the flexibility of the score driven weighted fitness model, we are able to take into account and explicitly model the strength of the links. We consider a link from bank $j$ to bank $i$ as present at week $t$ if bank $j$ lent money overnight to bank $i$, and the associated weight as the total amount lent over that period. This results in a set of $T = 298 $ weekly aggregated weighted networks. For a detailed description of the dataset, we refer the reader to \cite{barucca2018organization}.
\begin{figure}[h!]
\centering
\includegraphics[width=0.49\linewidth 	]{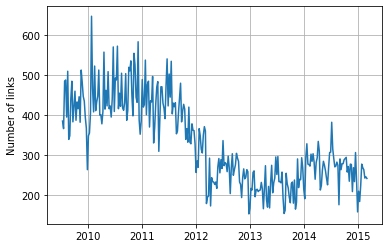}
\includegraphics[width=0.49\linewidth 	]{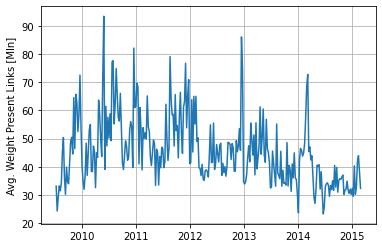}
\caption{Left panel: number of links present in the data at each time step. Right panel: average weight in Millions of Euro of the present links.}
\label{fig:emid_stats}
\end{figure}

As it is evident from the left panel of Figure \ref{fig:emid_stats}, the number of links in e-MID is significantly lower in the second half of the dataset. In particular it started declining in $2011$, most likely as a consequence of the European sovereign debt crisis and of the ECB unconventional measures, namely the long term refinancing operations (LTROs) that took place on the 22nd of December 2011 and on the 29th of February 2012 (see \cite{barucca2018organization} for an in-depth discussion). The number of links then fluctuated around a new lower level since the beginning of $2012$. As discussed in \cite{barucca2018organization}, the decreased number of links corresponds to a lower number of banks being active in the market. Both the network density~\footnote{Number of links present divided by the number of possible links, given the number of active banks.} and the average weight of present links has not followed a similar clear transition to a different level.

\subsection{Link and Weight Prediction}\label{sec:weight_pred}

As a first empirical application, we explore the possibility of using our approach to predict the presence of future links and the value of their weights. The problem of link prediction in temporal networks is extremely relevant in practical applications and has been discussed widely in the literature on binary networks \citep[][]{lu2011link, wang2015link, martinez2016survey, haghani2017systemic}. It can be defined in multiple ways depending on the context, the type of data at hand, and whether we want to predict the existence of a link in a partially observed network or the presence of a link in a future network. We focus on the latter case  that is often referred to as temporal link prediction. On the other hand, the prediction of weights in weighted temporal networks has received much less attention so far mainly due to the lack of models suited to describe both links and weights. For this reason, here we run an exercise using the full dataset described in the previous section and focus on forecasting the weights of the network at time $t+1$, using only a subset of the information available at time $t$. This means that,  when forecasting observations at time $t+1$, we use only observations from $t-T_{train}$ to $t$ for training, extremes included. This is very much in line with what we did in \cite{di2019score}, with the important difference that there we only considered binary link prediction, i.e. the sole prediction of links' presence. For a detailed discussion of how to employ the binary component of our model for the link prediction, we refer the reader to our previous work. 

Our first goal is to assess how effective can a model with time varying fitness in Eq. \eqref{eq:gen_gamma_fit_mod} be in forecasting the weights and to benchmark the proposed score driven approach against an alternative model. To this end, we compare the forecasts obtained using two methods, both based on Eq. \eqref{eq:gen_gamma_fit_mod} and differing only in their dynamics for the fitness. The first approach uses the score driven model to forecast the fitness at time $t+1$ with the update rule of Eq. \eqref{eq:fit_sd_update}. This is very easy to do in practice since, as mentioned in Section \ref{sec:new_mod_def}, the score driven fitness at time $t+1$ are deterministic conditionally on the observations at time $t$. The second approach is a combination of the sequence of single snapshot estimates, described at the beginning of Section \ref{sec:num_sim_dwfm}, and a set of AR(1) models, one for each fitness. In the latter approach, we first obtain the sequence of single snapshot estimates on the training set and then use them to estimate an AR(1) process for each fitness. Then, we use the estimated AR(1) models to forecast one value of the fitness at time $t+1$. Practically, we repeatedly estimate the two models on rolling windows of length $T_{train} = 100$ time steps and, once for each estimate, we forecast the first out-of-sample observation for each train window. We then compare the weights of present link at time $t+1$ with the expected values obtained from the two models and quantify the error by the mean squared error between the logarithms of observed and predicted weights 
$$
\text{MSE Log.} = \frac{\sum_{ij} \indfun\pt{\Yij\etp} \pt{\log\pt{\mathbb{E}\pq{\Yij | \widehat{\iwpar}\etp , \widehat{\iwpar}\etp} } -  \log \pt{\Yij\etp} }^2}{\sum_{ij} \indfun\pt{\Yij\etp}},
$$
where $\widehat{\iwpar}\etp $ and $ \widehat{\iwpar}\etp$ are the forecasts for the in and out weighted fitness, obtained using only observations up to time $t$. 
Similarly we compute the mean absolute difference (MAD)
$$
\text{MAD Log.} = \frac{\sum_{ij} \indfun\pt{\Yij\etp} \left\vert \log\pt{\mathbb{E}\pq{\Yij | \widehat{\iwpar}\etp , \widehat{\iwpar}\etp} } -  \log \pt{\Yij\etp} \right\vert }{\sum_{ij} \indfun\pt{\Yij\etp}}.
$$
We compare the logarithms of predicted and observed weights because the distribution of observed weights is quite heterogeneous. It roughly spans five order of magnitudes, and directly comparing the weights would result in a measure of goodness of fit mainly describing the fit of the largest weights~\footnote{Similar results hold using measures of relative error for the comparison between predicted and observed weights.}.

\begin{table}[ht]
\centering
\begin{tabular}{l|ccc}
	  Method \rule{0pt}{15pt}  & SD & SS - AR(1)  & Diebold-Mariano (p-value) \\
	\hline
	\rule{0pt}{12pt}  MSE Log. &  0.859  & 0.882 & $1.73 \times 10^{-7}$ \\
	\hline
	\rule{0pt}{12pt}  MAD Log. & 0.726 & 0.737 & $1.21 \times 10^{-7}$\\
 	\hline
\end{tabular}
\caption{ Weight prediction exercise: MSE and MAD between the logarithms of observed and predicted weights. First and second columns: results from the score driven (SD) model and the approach based on single snapshot  estimates and AR(1) processes ((SS) - AR(1)), respectively. Third column: p-values of a Diebold-Mariano test for the null hypothesis that the two forecasts are equivalent.} \label{tab:weights_prediction}  
\end{table}
From the results reported in Table \ref{tab:weights_prediction} we conclude that, similarly to the binary case, the score driven approach is a better choice to predict the weights with respect to a prediction based on sequence of single snapshot estimates, both in terms of MSE and MAD for the logarithms. The Diebold-Mariano \citep[][]{diebold2002comparing,harvey1997testing} test rejects the null hypothesis that the two forecasts are statistically equivalent.

\subsubsection{Comparison With Link Specific Regressions}

\begin{figure}[h!]
\centering
\includegraphics[width=0.9\linewidth 	]{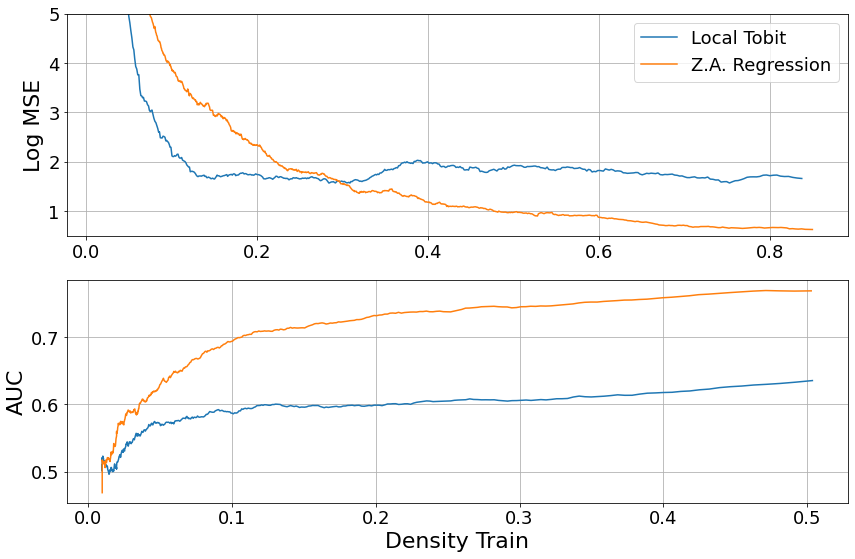}
\caption{ Out-of-sample Forecast accuracy as a function of the density in the training set for the models based on link specific regressions. Each line is obtained by averaging the log MSE (top panel) and the AUC (bottom panel) over rolling subsets of $1000$ links sorted by the fraction of non zero observations in the training sample. }
\label{fig:density_train}
\end{figure}

Among the models for weighted temporal networks reviewed in Section \ref{sec:lit_rev}, \cite{giraitis2016estimating} is the most relevant reference for the current work. 
Authors run a forecasting exercise on an interbank network of overnight loans. Differently from our case, they focus on the sterling interbank market that is very dense and somewhat small, when compared with the e-MID dataset. This is due to the structure of the UK interbank market that they consider. Indeed the fraction of links present at every time step for the full temporal network is always above $40\%$, while in our data it is typically equal to $6\%$. Furthermore, \cite{giraitis2016estimating} run a weight forecasting exercise restricted to the sub-network composed by the $4$ largest banks in the system, thus increasing the density of the actual considered network. They propose to model each link by means of a Tobit regression estimating the parameters with a local likelihood approach. Aside their methodological contributions, a key element of their work is the accurate selection of regressors to capture relevant network properties that are expected to influence the future network structure. They design a set of simple functions of the network at the previous time step and use them as regressors. Specifically, for the link going from bank $i$ to bank $j$, they consider as regressors the weight of that same link at the previous time step ({\it lagged}) plus the following quantities
\begin{itemize}
    \item lagged total daily amount lent by $i$ to all other banks except $j$ ;
    \item lagged total daily amount borrowed by $j$ from all other banks except $i$ ;
    \item lagged total daily amount lent by $j$ to all other banks except $i$ ;
    \item lagged total daily amount borrowed by $i$ from all other banks except $j$ ;
    \item lagged total daily lending and borrowing not involving either $i$ or $j$.
\end{itemize}
Here the goal is to compare their approach to link forecasting with ours, in an application to the e-MID data introduced in the previous sections, and investigate the reasons of different performances in links and weights prediction. In order to guide our intuition, in the following we also consider a version of their regression based on a simple Zero Augmented approach instead of the censoring that defines the Tobit approach. For the sake of simplicity, in this third approach, we separately consider a logistic regression on the binary part and a linear regression for the weights, both estimated with a standard, non local, maximum likelihood approach. 
As before we repeatedly estimate the models on rolling windows of length $T_{train}$ and report the results in Table \ref{tab:link_pred_regression_comparison}.
\begin{table}[]
    \centering
\begin{tabular}{l|rrrr}
    Model &  $T_{train}$ & MSE Log. &   MAD Log. &   AUC  \\
    \hline
    \midrule
    Localized Tobit      &      100 & 2.351 &  1.067 &  0.714  \\
    ZA Regression   &      100 & 2.785 &  1.240 &  0.830  \\
    Localized Tobit      &      200 & 2.267 &  1.043 &  0.795  \\
    ZA Regression   &      200 & 2.912 &  1.282 &  0.867  \\
    SD Generalized Fitness      &      100 & 0.859 &  0.726 &  0.896  \\
    \end{tabular}
    \caption{Results of the link and weights prediction exercise on e-MID data. We compare the Localized Tobit model of \cite{giraitis2016estimating} with a simple Zero Augmented regression that uses the same regressors. We also report the results of running the same exercise with the score driven generalized fitness model.}    
    \label{tab:link_pred_regression_comparison}
\end{table}
It emerges clearly that the score driven generalized fitness model achieves better results in forecasting both links and weights, for the dataset that we considered, with respect to the localized Tobit regression of \cite{giraitis2016estimating}. We believe this result to be mainly driven by two factors. The first one is that the Tobit model uses censoring to assign a finite probability to zero observations, i.e. missing links. Thus, the probability of observing a link and the expected values of the weights are modelled by the same set of parameters. In contrast, the generalized score driven fitness model employs zero augmentation to separately model the probability of observing a link and the expected weight of present links. Indeed, this intuition is confirmed by looking at the performances of the simple Zero Augmented regression, that significantly improves the prediction of the existence of links  -- as assessed by the area under the curve (AUC) measure -- but achieves slightly worst performances on predicting the weights of existing links with respect to the local likelihood Tobit. We empirically found the performances of both models to deteriorate significantly for the links that are rarely observed in the training set.  In fact, as we show in Figure \ref{fig:density_train}, both the out-of-sample log-MSE and AUC deteriorate when computed for subsets of links with a low fraction of non zero observations in the training set. Interestingly, the log MSE of the ZA regression seems to be monotonically decreasing when the density in the training set increases, while the Tobit regression does not. We interpret this as an additional indication of the advantages in separating the modelling of links' presence from their expected weight.

The second factor is that approaches based on running one regression for each link are extremely over-parametrized. They require the estimation of a large number of parameters that scales with the number of links ($N\pt{N -1}$), as opposed to our approach which requires the estimation of $6$ parameters per each node, thus scaling as $N$. While the idea of running a separate model for each link has clear advantages for parallelizing the execution, such a high number of parameters can result in overfit and poor out-of-sample generalization, when the dataset at hand does not allow for a large number of time steps to be used in training. Indeed we corroborate this intuition by noting that the results of both the localized Tobit and the Zero Augmented regression improve if we consider a training set of $200$ time steps, instead of one of $100$ (see the third and fourth lines of Table \ref{tab:link_pred_regression_comparison}).

\subsection{The Effect of Interest Rates on Interbank Lending}\label{sec:app_eonia}

In this section, by means of the score driven model, we investigate the effect of interest rates on the dynamics of the interbank network data introduced above. To track average interest rates, we use the EONIA benchmark. EONIA ``\textit{is a measure of the effective interest rate prevailing in the Euro interbank overnight market. It is computed as a weighted average of the interest rates on unsecured overnight contracts on deposits denominated in Euro, as reported by a panel of contributing banks}"~\footnote{Definition from \url{https://stats.oecd.org/}.}.

Intuitively we expect that banks' funding rates and the topology of the interbank market are deeply related. This relation is of clear interest from the point of view of policymakers and has received much attention in the literature \citep[see, for instance,][]{akram2010interbank, iori2015bank, arciero2016measure, temizsoy2017network, brunetti2019interconnectedness}.  Of particular relevance for the results discussed in this section is the work of  \cite{akram2010interbank} that investigated the effects of banks' characteristics and the conditions of the network as a whole on the interest rates that each bank faces on the interbank market. They exploited a remarkable dataset, obtained from Norges Bank real time gross settlement system, that allowed them to model bank specific interest rates as dependent on a set of variables and controls, including overall market's liquidity. From basic supply and demand reasoning one would expect that excess liquidity in the market would have a negative pressure on the interest rates on average, and indeed they found that interest rates tend to be lower when the overall liquidity available on the market is higher.

A second relevant work for the purposes of this section is \cite{brunetti2019interconnectedness}, where authors considered data on the e-MID interbank market for a period ranging from the beginning of 2006 to the end of 2012, focusing solely on the binary part of the daily temporal network of overnight loans. They computed various aggregated network statistics for each time step, thus obtaining one univariate time series for each statistic. Among other quantities, they computed the density of each network -- defined as the number of connections as a proportion of all possible connections -- and, using a standard linear regression, found it to be positively related with EONIA.

In the following, we explore the impact of interest rates on the probability of observing each link and on the expected weight of observed links by applying the score driven weighted fitness model to the e-MID dataset, using as external covariate the EONIA rate. For our estimates, we use a training set comprising the first $80\%$ of time steps and left the last $20\%$ to assess goodness of fit out-of-sample. Moreover, similarly to what done in the numerical simulations discussed in Section \ref{sec:num_sim_dwfm}, we compare the score driven time varying fitness with two alternative specifications: a model without fitness and one with constant fitness, as defined in Section \ref{sec:num_sim_dwfm}, both with EONIA as the only external variable. We then compare their goodness of fit both in-sample and out-of-sample. In Table \ref{tab:emid_eonia_gof} we show the results that clearly confirm the importance of including time varying fitness to improve goodness of fit.  We report the Bayesian Information Criterion (BIC) for each model, computed separately for the likelihood of observing links and the likelihood of their weights, to compare goodness of fit in-sample. The binary and weighted parts of each model are evaluated separately out of sample. We quantify out of sample accuracy in predicting link's presence by means of the AUC, while for the weights we compute the MSE of the logs of the weights, only for the present links. 
\begin{table}[ht]
\centering
\begin{tabular}{l|ccc}
	Model \rule{0pt}{15pt}  &    No Fitness &    Constant Fitness   &   SD Fitness \\
	\hline
	\hline
\rule{0pt}{12pt}  BIC Bin & $1.75 \times 10^6$& $0.53\times 10^6$  & $0.45 \times 10^6$ \\
	\hline
	\rule{0pt}{12pt}  BIC Weight & $1.90 \times 10^6$   & $1.46 \times 10^6$ & $1.46 \times 10^6$ \\
	\hline
	\rule{0pt}{12pt}  AUC - Test Set &0.48 & 0.82  & 0.92 \\
	\hline
	\rule{0pt}{12pt}  MSE Log. - Test Set & 58.15 & 1.01  & 0.78 \\
\end{tabular}
\caption{EONIA effect on e-MID: in sample BIC, for both the binary and weighted part of model (\ref{eq:gen_gamma_fit_mod}), AUC for the out of sample evaluation of the binary part, and the MSE of the logarithms for the out of sample evaluation of the weighted part. First and second columns: results for a model without and with constant fitness, respectively. Third column: score driven fitness.} \label{tab:emid_eonia_gof}  
\end{table}
Since the model without fitness is clearly not a good fit for the data we do not discuss it further, and in Table \ref{tab:emid_eonia_par} we report the estimated regression coefficients using the models with constant and score driven fitness.
\begin{table}[ht]
\centering
\begin{tabular}{l|c|r}
	Model \rule{0pt}{15pt}  &  Constant Fitness   &   SD Fitness \\
	\hline
	\hline
 	\rule{0pt}{12pt}  $\beta_{bin}$  & $0.69\pm 0.06$& $0.29\pm 0.05$\\
 	\hline
 	\rule{0pt}{12pt}  $\beta_{w}$   & $0.022\pm 0.029$  & $-0.13 \pm 0.02$ \\
\end{tabular}
\caption{EONIA effect on e-MID. Estimates of the regression coefficients using models with constant or score driven fitness.} \label{tab:emid_eonia_par}  
\end{table}
The model with score driven fitness is clearly the best fit for the data, both in sample and out of sample, as measured by the metrics reported in Table \ref{tab:emid_eonia_gof}. Moreover, the parameters estimated by the score driven fitness model are always statistically significant while the $\beta_w$ estimated from the constant fitness model is not. We interpret this discrepancy as a sign that disregarding the fitness dynamics can lead us also to miss-guided qualitative interpretations, consistently with the numerical results discussed in Section \ref{sec:num_sim_dwfm_unobs}. 

From the estimate $\beta_{bin}  = 0.29\pm 0.05$, we can deduce that, in the considered period, the probability of observing a link in the network is positively related with the interest rates, hence the lowering of interest rates tends to reduce the overall market interconnectdness, even taking into account  bank specific effects captured by the fitness. This result is coherent with the relation between network density and EONIA found in \cite{brunetti2019interconnectedness}, although the approach based on standard regression on aggregated network statistics is different from ours. In fact, thanks to the time varying binary fitness in our model, the estimated effect of EONIA is decoupled from bank specific effects that are instead accounted for by the fitness. Such a separation of bank specific effects from the impact of a covariate is instead not possible when considering the density of the whole network as done in \cite{brunetti2019interconnectedness}. 


For what concerns the effect on the liquidity exchanged through the observed links, i.e. the links' weights, the estimated $\beta_{w} = -0.13 \pm 0.02$, indicates that the weight of the observed overnight loans is negatively related with the average interest rate in the market. Our result is coherent with the work~\cite{akram2010interbank} on the Norwegian interbank market, but our methodology allows us to explore a different aspect of the relation between liquidity and interest rates. In fact, that result regards the relation between bank specific rates and aggregated liquidity, while we explore the relation between average rates on the market and the weight of present links, controlling for time varying bank specific effects by means of the time varying fitness. Thanks to zero augmentation and separate modelling of links and weights, our finding is directly related with the average magnitude of the overnight loans that are actually present, more than with the total liquidity in the market. In summary, the data considered indicate that lower interest rates are related with a reduction of network interconnectdness but an increase of the average liquidity flow for the loans that are present.

Finally, we mention that, as we show in Appendix \ref{sec:filtered_fitness_dyn}, if we do not include EONIA as external covariate the dynamical fitness tend to correlate with it. This corroborates the fact that the estimated coefficients are found to be statistically significant.

We point out that our results on the relation between the dynamics of the e-MID interbank network and EONIA are obtained leveraging the full information available in the description of a temporal network as a temporal sequence of matrices, and considering the impact of the covariate both on the probability of each individual link and the expected weight of observed links. Differently from \cite{akram2010interbank} and 
\cite{brunetti2019interconnectedness}, we do not need to collapse the matrices into a single network statistic to estimate the effects of external variables. We directly use matrix valued network data and, thanks to the time varying latent fitness parameters, we can decouple the impact of EONIA from unobserved time varying node specific effects. The advantage of using matrix valued network data will become even more evident in the next section where we consider link specific covariates and carry out an analysis that would be impossible with standard regression methods on univariate network statistics.

\subsection{Link and Weight Persistence}

As a final application, we use our model to contribute to the literature on the persistence in interbank networks \citep[][]{weisbuch2000market, cocco2009lending, hatzopoulos2015quantifying, mazzarisi2020dynamic} by exploring both the persistence of links and that of the weights. The existence of privileged lending relations between pairs of banks is a well known phenomenon and it is often referred to as \textit{preferential trading} \citep{weisbuch2000market}. The motivations behind it can be explained by the relevance of strong lending relationships between banks as a way to overcome monitoring of creditworthiness and limit the risk of counter-party default \citep{cocco2009lending}. The existence of preferential trading behaviours has been assessed quantitatively by means of statistical methods specifically developed for the purpose \citep{hatzopoulos2015quantifying}, in the case of binary networks. Additionally, models for binary temporal networks  have been proposed that explicitly take it into account \citep{mazzarisi2020dynamic}. 

In this section we exploit the flexibility of our model and estimate the effect of two predetermined covariates that are meant to capture the persistence of links and weights. 
For what concerns link persistence of the binary network, we explore how the presence of a link at time $t-1$ influences the probability of observing a link at time $t$. That amounts to use $\indfun\pt{\Yijtm}$  as covariate in Eq. \eqref{eq:gen_fit_bin_prob}.
To assess persistence in the links' weights, we estimate the effect of the weight of a link at $t-1$ in determining its weight at time $t$ by using $\log \pt{\Yijtm}$ as covariate in Eq. \eqref{eq:gen_fit_cond_exp}. Let us recall that we have tested numerically the possibility to estimate such effects in synthetically generated data in Section \ref{sec:num_sim_dwfm}.
\begin{table}[ht]
\centering
\begin{tabular}{l|ccc}
	Model \rule{0pt}{15pt}  &    No Fitness &    Constant Fitness   &   SD Fitness \\
 	\hline
 	\hline
 	\rule{0pt}{12pt}  $\beta_{bin}$ & $0.064 \pm 0.038$& $2.875\pm 0.045$ &  $2.048 \pm 0.045$ \\
 	\hline
 	\rule{0pt}{12pt}  $\beta_{w}$ & $1.050 \pm 0.003 $ & $0.073\pm 0.002$& $0.064\pm 0.002$ \\
	\hline
	\hline
\rule{0pt}{12pt}  BIC Bin & $5.66 \times 10^6$& $4.46\times 10^6$  & $4.16 \times 10^6$ \\
	\hline
	\rule{0pt}{12pt}  BIC Weight & $2.61 \times 10^6$   & $1.45 \times 10^6$ & $1.45 \times 10^6$ \\
	\hline
	\rule{0pt}{12pt}  AUC - Test Set &0.748 & 0.882  & 0.932 \\
	\hline
	\rule{0pt}{12pt}  MSE Log. - Test Set & 21.22 & 0.90  & 0.76\\
\end{tabular}
\caption{ Results on estimates of link persistence in e-MID. One column for each one of the three alternative model specifications. 
In the third and fourth rows, we show the in sample BIC for the binary and weighted parts of the model in \eqref{eq:gen_gamma_fit_mod}, respectively. The last two rows are out of sample measures of goodness of fit. The fifth one is the out of sample AUC for the binary part. The last row is the out of sample MSE of the logarithms of the weights.}  \label{tab:emid_persist}  
\end{table}
As in the previous section, we compare three models, a model without fitness, one with constant fitness, and the score driven fitness model, all using the same external covariates and two scalar coefficients, $\beta_{bin}$ and $\beta_w$ that quantify the persistence of links and weights respectively. 
The results in Table \ref{tab:emid_persist} confirm that neglecting node specific time varying effects results in worst fitting of the data. This is evident by looking at the superior performances, both in-sample and out-of-sample of the models with score driven fitness, with respect to those without or with constant fitness. 
The three model specifications all result in positive coefficients both for the binary and the weighted covariates. With the best performing model among those three, the model with score driven fitness, we estimate $\beta_{bin} = 2.048 \pm 0.045$. This indicates that globally the presence of a link at time $t-1$ positively impacts the probability of observing that same link at time $t$. This is in agreement with the current consensus in the literature, supporting preferential trading behaviours in e-MID, that has been validated empirically only on the binary part of temporal interbank networks, for example by \cite{hatzopoulos2015quantifying} and \cite{mazzarisi2020dynamic}. The novel aspect of our analysis lays in the estimated $\beta_{w} = 0.064 \pm 0.002$, that highlights a weight persistence effect. This result complements the analysis of \cite{hatzopoulos2015quantifying}, as they considered the weighted networks of the number of loans between each pair of banks, neglecting altogether the amount lent for each loan. By design, our model allows us to highlight the tendency of banks to form links whose weight is positively related with the weight at previous steps, a tendency that we might refer to as weight persistence.

\section{Conclusions}\label{sec:conclusions}

In this work, we proposed a model for the description of sparse and weighted temporal networks that extends the well known fitness model for static binary networks. In the new score driven weighted fitness model, we also model links' weights with an additional set of fitnesses. Both binary and weighted fitness follow a stochastic dynamics driven by the score of the conditional likelihood. Additionally, we also considered the possibility for the network dynamics to depend on a set of external covariates. 

Our numerical simulations proved the advantages of the score driven fitness over static fitness and over a sequence of standard cross sectional estimates. As an empirical application, we investigated the determinants of the dynamics of links and weights in the e-MID interbank network. We proved that there is a significant advantage in using score driven time varying parameters to forecast weights, with respect to single snapshot estimates. We exploited the flexibility of the new model to estimate the impact of the EONIA rate in determining the links and weights dynamics. We used it to inform the discussion on persistence in interbank networks providing empirical evidence of weights' persistence.

We run an empirical analysis on the prediction of links and weights that highlighted the superior performances of our approach with respect to alternatives based on single snapshot estimation and link specific regressions. Most notably, for the dataset that we considered,  we found our model to attain clearly superior performances with respect to the local likelihood Tobit model by \cite{giraitis2016estimating}. In short, we believe this to be a direct consequence of our modelling choices that allow for a flexible description with a moderate number of parameters. Specifically, using zero augmentation we describe separately the probability of a link to exist and its expected weight, and leveraging the score driven approach we only need to estimate order $N$ parameters instead of $N^2$.

Our work provides several perspectives for future research. First, the possibility to jointly model and predict the presence of a link and the associated weight could find relevant applications in the financial stability literature. Weighted financial networks are known to be among the determinants of systemic risk and their dynamical description has so far neglected the role of the weights. Second, score driven weighted fitness model could be applied on multiple instances of real world sparse and weighted temporal networks, where the standard approach of ignoring the weights might result in significant information loss.  Finally, we plan to investigate the temporal evolution of the community structure of real world networks. Community detection has attracted an enormous amount of attention in various streams of literature \citep{javed2018community}, in particular in the context of temporal networks \citep{rossetti2018community}. We believe that the score driven fitness approach could provide a flexible modeling framework and offer valuable support to assess the degree of persistence of a given partition of nodes into groups. 
\newpage
\bibliography{sparse_weighted_nets}

\appendix

\section{Alternative Distributions for the Weights}\label{sec:weight_distrib_SI}
In the main text, we employed the gamma distribution to model the links' weights. It is extremely easy to move to different specifications. If, for example, we consider the log-normal distribution, we can substitute its parametric form
\begin{equation*}
g\uij\tonde{y} = \frac{1}{y\sigma {\sqrt {2\pi }}} e^{\frac{- \tonde{\ln y-\mu_{ij}}^{2}}{2\sigma^{2}}}
\end{equation*}
in place of the gamma distribution, and obtain a different update equation for the time varying parameters. The fitness would be driven by the score of the log-normal
\begin{equation*}
 \nabla_i\pt{\Yt, \iwpar, \owpar,\sigma}
 = \sum_j \pt{\frac{1}{\sigma^2} \log\frac{\mathbb{E}\quadre{Y\uij\et\vert Y\uij\et>0}}{Y\uij\et}   -   \indfun\pt{\Yij} }.
\end{equation*}
Indeed, extending our approach to different specifications is practically straightforward. The interested reader would only need to compute the log-likelihood, sample from the distribution, and compute the score and scaling matrix. Conveniently the log-likelihood and sampling methods are already available for many distributions in PyTorch \citep{NEURIPS2019_9015}, upon which we built the core library for this work available at \url{https://github.com/domenicodigangi/dynwgraphs}.   

\section{Fitness Identification}\label{sec:identification_SI}
It is easy to see that, as for the static version, the fitness parameters of the fitness model for a directed network are not identified. This fact is well known for the binary fitness model \citep[][]{yan2016asymptotics} and remains true for the weighted fitness that we consider here. Let us indicate here with $\ibwpar$ and $\obwpar$ the vectors of in and out fitness respectively, without specifying whether they are binary or weighted. With this notation it is immediately clear that the following transformation 
\begin{align*}\label{eq:id_issue_stat_fit}
\ibwpar_i \rightarrow \ibwpar_i + c, \quad \forall i \\ 
\obwpar_i \rightarrow \obwpar_i - c, \quad \forall i
\end{align*}  does not change the sum $\ibwpar_{i} + \obwpar_{j} + \beta \Xij$ and leaves the probability distribution unchanged, for any $c\in {\mathbb R}$. The parameters are not uniquely identified and, in order to compare estimates across different time steps, we need to impose an identification restriction, as commonly done in the literature \citep{yan2016asymptotics, mazzarisi2020dynamic}. In this work we require that 
\begin{equation*}
\sum_i \ibwpar_i = \sum_j \obwpar_j, 
\end{equation*}
at each time step.

\section{Filtered Fitness Dynamics Wwith External Covariates}\label{sec:filtered_fitness_dyn}

Here, we inspect the dynamical behaviour and the role of the fitness filtered with the score driven weighted fitness model, defined in Section \ref{sec:new_mod_def}, in presence of external covariates. In particular, we discuss how the dynamical evolution of the filtered fitness changes when we consider the dependency on EONIA, estimated as described in Section \ref{sec:app_eonia}, with respect to the case of no covariate.

\begin{figure}[h!]
\centering
\includegraphics[width=0.8\linewidth 	]{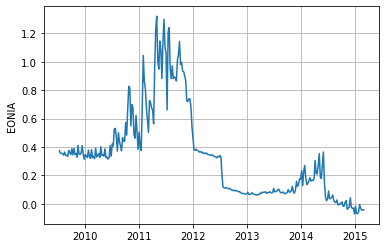}
\caption{Weekly average of EONIA rate over the considered time period.}\label{fig:eonia_rate}
\end{figure}

To this end, for a subset of the most active links~\footnote{We consider the links that are present at least  $5\%$ of the weeks.}, we compute the sum of the in and out fitness corresponding to their probabilities, and expected weights. For example, for link $\pt{i,j}$ we consider $\widehat{\ibinpar}_i + \widehat{\obinpar}_j$ and $\widehat{\iwpar}_i + \widehat{\owpar}_j$. 
We compute these quantities using two sets of filtered fitness. The first set is obtained by estimating and running the score driven weighted fitness model as a filter without considering any external covariate, similarly to what we did in Section \ref{tab:weights_prediction}. The second set is obtained with a similar approach, now including in the model specification also the dependency on EONIA, exactly as in Section \ref{sec:app_eonia}. Then, given these two sets of fitness sums  we look at their Spearman rank correlation \citep[][]{spearman1961proof} over time with EONIA, thus obtaining four correlations for each considered link: for each of the two sets of filtered fitness (with and without EONIA as covariate) we have one correlation for the binary fitness and one for the weighted fitness. We consider the sum of fitness because, as it is evident from \eqref{eq:gen_fit_bin_prob} and \eqref{eq:gen_fit_cond_exp}, they are related to the probability of observing a link and to the expected weight of that link. Moreover, given that the EONIA rate had a substantial drop around the middle of $2012$, as showed in Figure \ref{fig:eonia_rate}, in computing the  correlation with the sums of fitness we split the data set in two periods and consider them separately.

\begin{figure}[!ht]
\centering
\includegraphics[width=\linewidth 	]{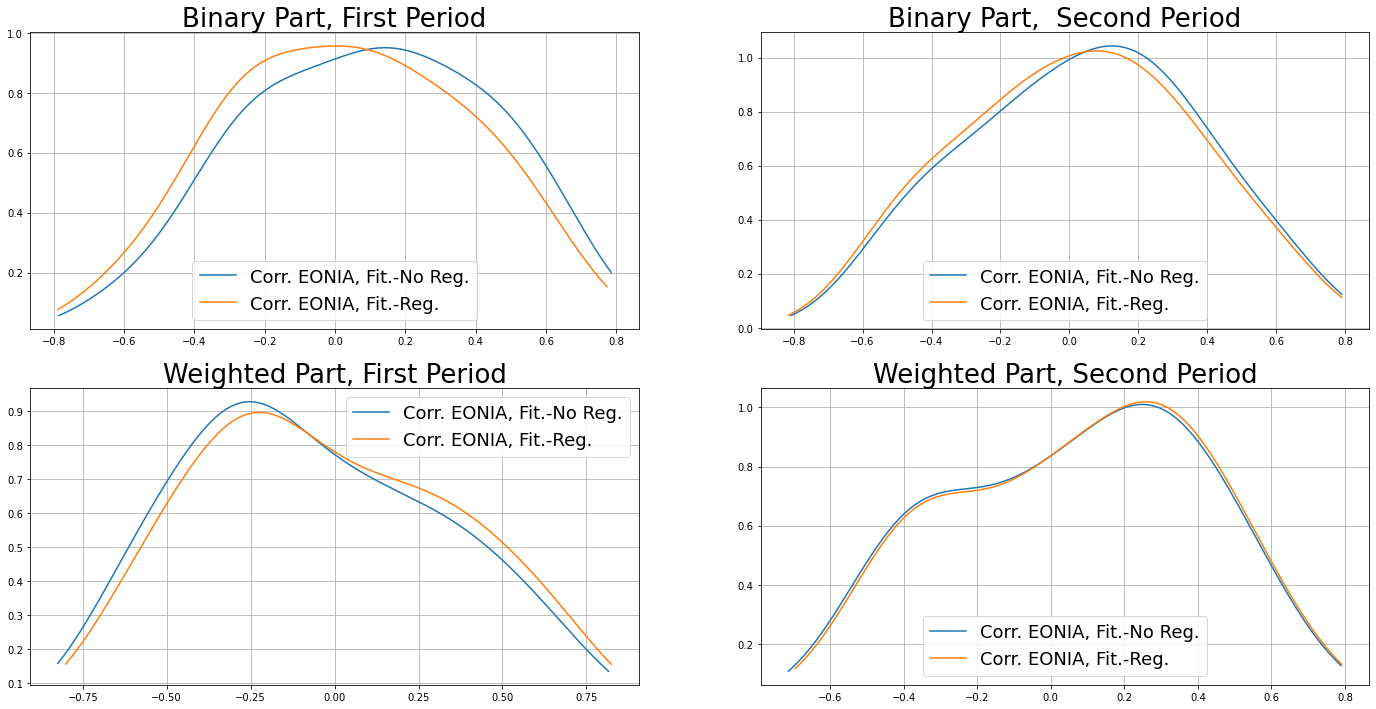}
\caption{Distribution of the Spearman correlation between the fitness sums and EONIA computed in two periods. The first period ranges from the beginning of the sample up to $15/07/2012$, the second one from $22/07/2012$ to the end of the sample. Left panels: correlations computed for the first period; right panels: correlations computed in the last period. Top panels: correlations between EONIA and the sums of binary fitness; bottom panels: correlations with the weighted fitness. }\label{fig:fit_sum_eonia_corr}
\end{figure}

The plots shown in Figure \ref{fig:fit_sum_eonia_corr} highlight the different behaviour of the fitness, with respect to EONIA, when the latter is explicitly considered as an external covariate. We notice that in the first period, i.e. when EONIA rates are higher, the correlation between the fitness sums and EONIA are significantly different when we include EONIA as covariate or we do not. To statistically confirm the difference we run a Kolmogorow-Smirnov test having as null hypothesis the fact that the two distributions are identical. In the first period, the null hypothesis is rejected at $5\%$ confidence level for both the binary and weighted part. In the second period instead, we cannot reject the hypotheses that the two distributions are the same.  Hence the relation of the fitness with EONIA in the second period does not seem to be affected by whether the latter is explicitly included or not as external covariate. 

We believe that the tendency of the dynamical fitness to correlate with EONIA, when the latter is not considered explicitly as an external variable, corroborates the fact that the estimated coefficients are found to be significantly different from zero.

\begin{figure}[!ht]
\includegraphics[width=\linewidth 	]{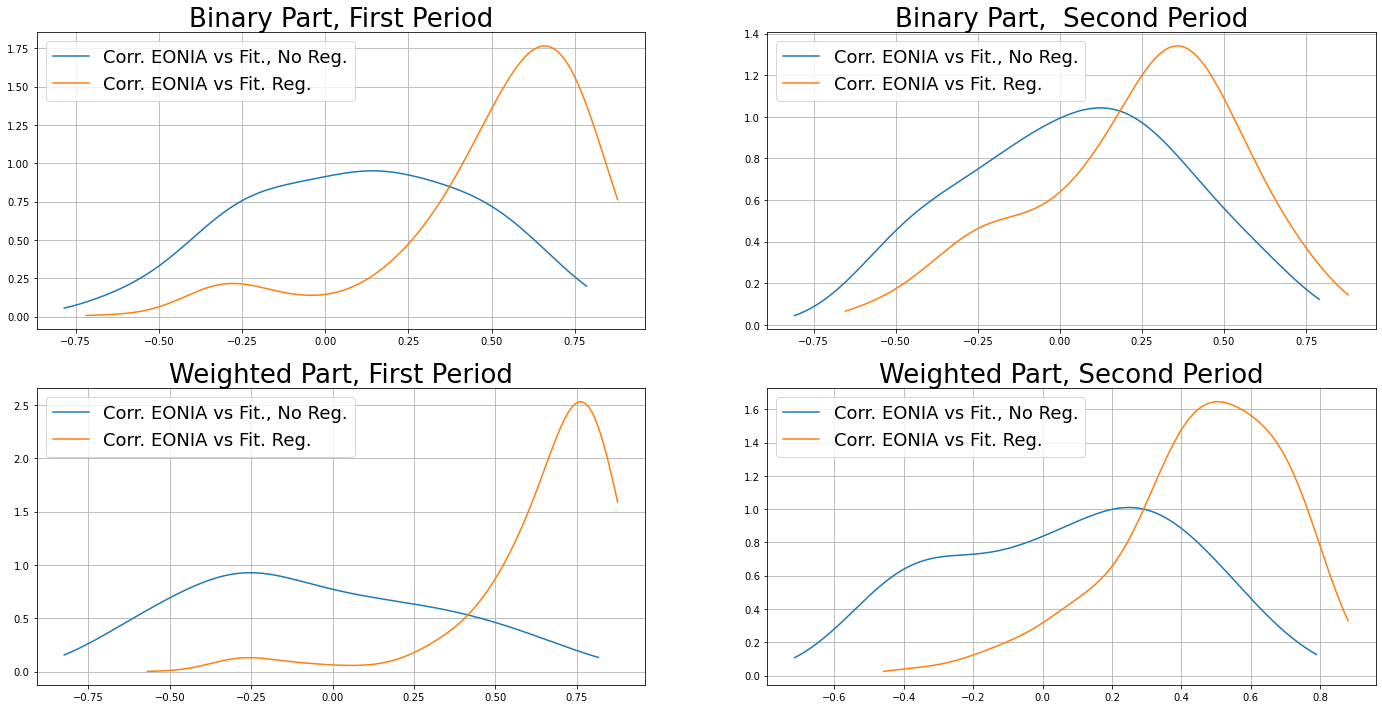}
\caption{Similar plots as in Figure \ref{fig:fit_sum_eonia_corr}, now repeated after forcing the values of the regression coefficients to $\beta_{bin} = -3$ and $\beta_{w} = -3$.
}\label{fig:fit_sum_eonia_corr_fake_neg_reg}
\end{figure}

\begin{figure}[!ht]
\includegraphics[width=\linewidth 	]{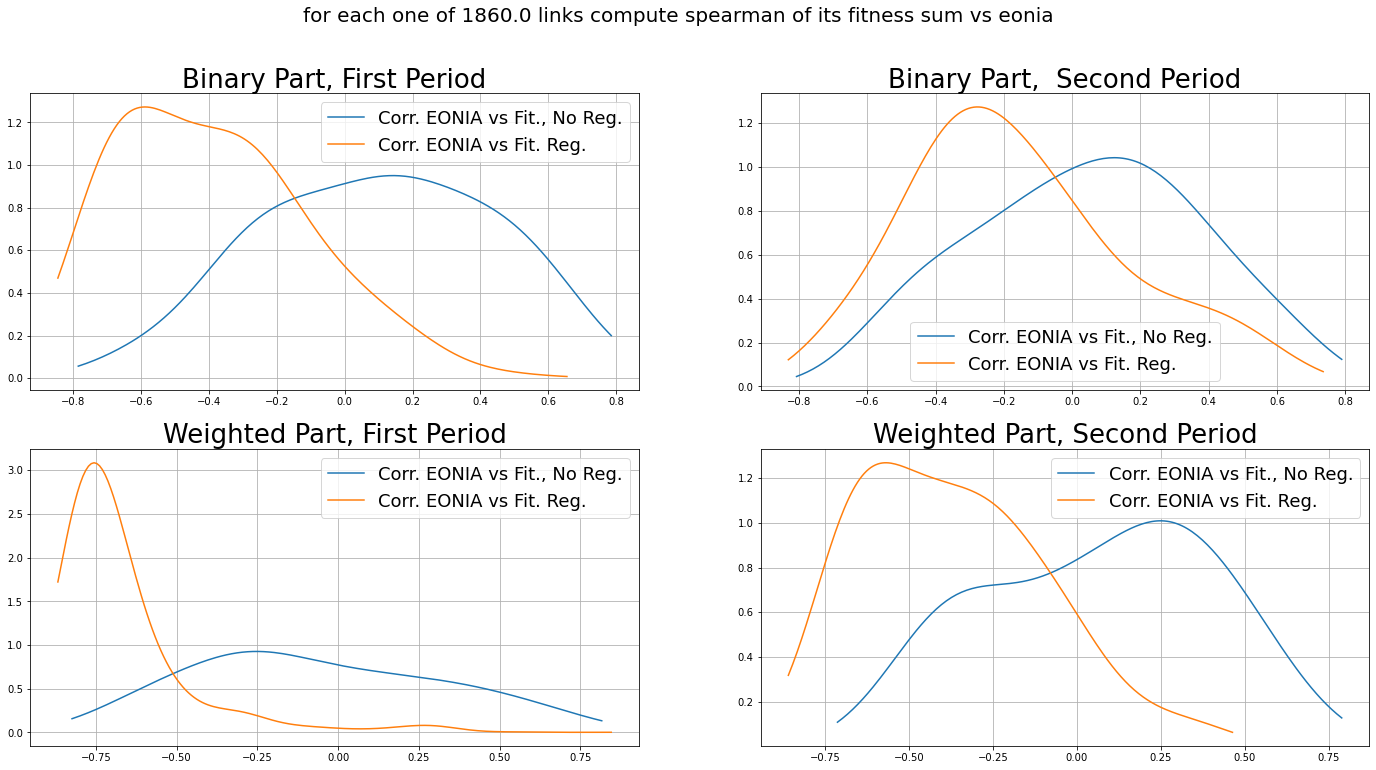}
\caption{Similar plots as in Figure \ref{fig:fit_sum_eonia_corr}, now repeated after forcing the values of the regression coefficients to $\beta_{bin} = 3$ and $\beta_{w} = 3$.
}\label{fig:fit_sum_eonia_corr_fake_pos_reg}
\end{figure}

In order to better understand this behaviour we repeated the same analysis artificially modifying the regression coefficients before filtering the fitness. We keep everything equal except the values of $\beta_{bin}$ and $\beta_w$. Instead of their MLE values, we now artificially set the values to $\beta_{bin} = 3$ and $\beta_w = 3$ in Figure \ref{fig:fit_sum_eonia_corr_fake_pos_reg} and $\beta_{bin} = -3$ and $\beta_w = -3$ in Figure \ref{fig:fit_sum_eonia_corr_fake_neg_reg}. 
In practice, we change the regression coefficients and then filter the time varying fitness using the score driven update rule. We notice that, when the regression coefficient is artificially inflated, see Figure \ref{fig:fit_sum_eonia_corr_fake_pos_reg}, the fitness sums tend to negatively correlate with EONIA. While, when we force the coefficients to be negative, the fitness positively correlate with EONIA. In both cases the filtered fitness behaviour tends to mitigate the impact of the spurious regression coefficients. These effect is more evident in the first period than in the second, as expected due to the higher values of EONIA.

\end{document}